# ACQUISITION PROBABILITY OF MULTI-USER UWB SYSTEMS IN THE PRESENCE OF A NOVEL SYNCHRONIZATION APPROACH

Moez Hizem[1] and Ridha Bouallegue[1]

[1]Innov'Com Laboratory, Sup'Com, University of Carthage, Tunis, Tunisia

```
moezhizem@yahoo.fr
ridha.bouallegue@supcom.rnu.tn
```

## ABSTRACT

*In this paper, to synchronize Ultra Wideband (UWB) systems in ad-hoc multi-user environments, we propose a new timing acquisition approach for achieving a good performance despite the difficulties to get there. Synchronization constraints are caused by the ultra-short emitted waveforms nature of UWB signals. Used in [1, 2] for single-user environments, our timing acquisition approach is based on two successive stages or floors. Extended for multi-user environments, the used algorithm is a combination between coarse synchronization based on timing with dirty templates (TDT) acquisition scheme and a new fine synchronization scheme developed in [3-6] which conduct to an improved estimate of timing offset. In this work, we develop and test this method in both data-aided (DA) and non-data-aided (NDA) modes. Simulation results and comparisons are also given to confirm performance improvement of our approach (in terms of mean square error and acquisition probability) compared to the original TDT algorithm in multi-user environments, especially in the NDA mode.*

## 1. INTRODUCTION

UWB systems have received a great attention in the last years. Furthermore, the interest for commercial UWB technology is growing fast especially in the areas of high-data rate short-range wireless multimedia applications, as well as for low-data rate sensor networks [7]. One of these benefits comes from the large number of users allowed access with Time Hopping (TH) codes, and potential to overlay existing narrowband systems such as IEEE 802.11 and Bluetooth [8]. Hence, the idea of this work is to extend the timing acquisition approach previously developed in [1, 2] for multi-user environments this time.

Nevertheless, to exploit these benefits, one of the most important challenges (at least at the physical layer) is to obtain an accurate timing synchronization and more specifically timing offset estimation. In general, synchronization is usually achieved in two stages [9]. The first stage realizes coarse synchronization to within a realistic amount of precision in a short time, and is well-known as the acquisition stage. For this, we use the Timing with Dirty Templates (TDT) acquisition scheme introduced in [10] and developed in [11-14]. The TDT approach is an attractive technique for UWB systems, which is specified d by its low complexity and rapid acquisition in the DA mode. It's based on correlating adjacent symbol-long segments of the received waveform.

To improve the acquisition probability and the performance of multi-user UWB systems, we propose in the second stage a novel fine synchronization approach known as tracking stage which is responsible for preserving synchronization through clock drifts tacking place in the transmitter and the receiver. Tracking is usually accomplished with a delay locked loop (DLL) [9]. Timing acquisition is an enormous difficulty caused by UWB systems. Compared with the original TDT synchronizer, simulation results show that our new based-TDT-synchronizer can achieve a higher acquisition probability than the original TDT in both NDA and DA modes for multi-user environments.

The rest of this paper is organized as follows. The ensuing Section 2 describes the UWB TH-PAM system in multi-user environments. The Section 3 outlines our novel acquisition algorithm based on two stages. In Section 4, the simulations are carried out to corroborate our analysis in comparison with the original TDT approach. And finally, the conclusions are given in Section 5.

## 2. SYSTEM MODEL FOR MULTI-USER LINKS

The UWB time hopping impulse radio signal considered in this paper is a stream of narrow pulses, which are shifted in amplitude modulated (PAM). The same modulated pulse is repeated $N_f$ times (frames number) over a $T_s$ period (symbol time). During each duration frame $T_f$, a data-modulated ultra-short pulse p(t), with duration $T_p \ll T_f$, is transmitted [4]. The transmitted waveform from the *u*th user is

$$v_u(t) = \sqrt{\varepsilon_u} \sum_{k=0}^{+\infty} s_u(k) p_{u,T}(t - kT_s) \quad (1)$$

where $\varepsilon_u$ represents the energy per pulse, $s_u(k)$ are differentially encoded symbols and drawn equiprobably from finite alphabet. In our case, $s_u(k)$ symbolize the binary PAM information symbols and $p_{u,T}(t)$ indicates the transmitted symbol as

$$p_{u,T}(t) := \sum_{i=0}^{N_f - 1} p(t - iT_f - c_u(i)T_c) \quad (2)$$

where $T_c$ is the chip duration and $c_u(i)$ is the user-specific pseudo-random TH code during the *i*th frame.

The transmitted signal propagates through the multipath channel corresponding to each user. The UWB channel is modelled as tapped-delay line with $L_u$ taps, where $\{\alpha_{u,l}\}_{l=0}^{L_u-1}$ and $\{\tau_{u,l}\}_{l=0}^{L_u-1}$ is amplitude and delay of the L multipath elements, respectively. The channel is assumed quasi-static and among $\{\tau_{u,l}\}_{l=0}^{L_u-1}$, $\tau_0$ represents the propagation delay of the channel. Thus, the received waveform from all users is

$$r(t) = \sum_{u=0}^{N_u - 1} \sqrt{\varepsilon_u} \sum_{l=0}^{L_u - 1} \alpha_{u,l} v_u(t - \tau_{u,l} - \tau_u) + \eta(t) \quad (3)$$

where $N_u$ is the users number, $\tau_u$ is the propagation delay of the *u*th user's direct path and $\eta(t)$ is the zero-mean additive Gaussian noise (AGN). The global received symbol-long waveform is therefore given by

$$p_{u,R}(t) := \sum_{l=0}^{L_u - 1} \alpha_{u,l} p_{u,T}(t - \tau_{u,l}) \quad (4)$$

Assuming that the nonzero support of waveform $p_{u,R}(t)$ is upper bounded by the symbol time $T_s$, the received waveform in (3) can be rewritten as

$$r(t) = \sum_{u=0}^{N_u-1} \sqrt{\varepsilon_u} \sum_{k=0}^{+\infty} s_u(k) p_{u,R}(t - kT_o - \tau_u) + \eta(t) \quad (5)$$

In the next section, we will develop a low-complexity timing acquisition approach using TDT synchronizer in order to find the desired timing offset for multi-user environments. The structure of our synchronization scheme is illustrated in Fig.1.

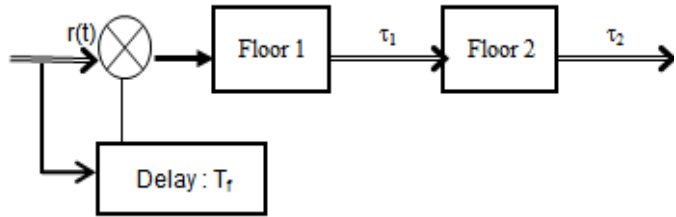

Figure 1. Structure of our synchronization scheme

## 3. PROPOSED TIMING ACQUISITION APPROACH

There are many difficulties reaching from the signal and channel characteristics relied to UWB systems that show the importance of the timing acquisition problem and the necessity to resolve it effectively, in particular for multi-user environments. This signifies that there could be numerous steps in the research domain which may be judged satisfactory and may be developed to accelerate the timing acquisition processing.

As mentioned previously, the new acquisition timing scheme proposed in this paper for multi-user environments consists of two complementary stages or floors. The first one is based on a blind (or coarse) synchronization algorithm which id the well-known TDT. The system model structure's with first stage synchronization is shown in Fig.2. This approach is consisted on correlating adjacent symbol-long segments of the received waveform. The basic idea behind TDT is trying to find the maximum of square correlation between pairs of successive symbol-long segments. These symbol-long segments are also called dirty templates and they are subject to the unknown offset $\tau_0$. Then, we will analyze $\tilde{\tau}_0$ representing estimate offset of $\tau_0$ by deriving upper bounds on their mean square error (MSE) in both NDA and DA modes.

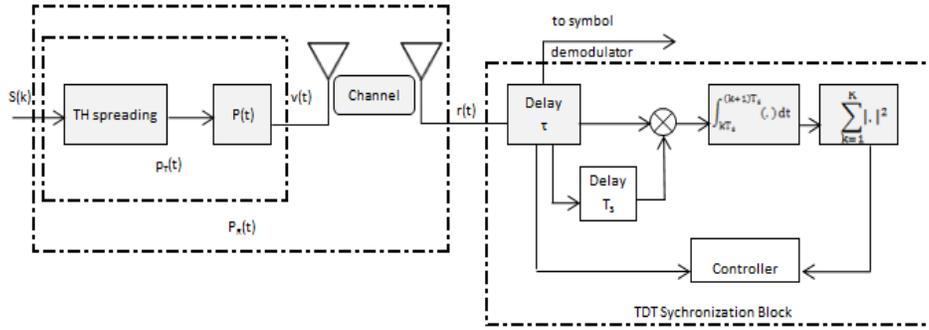

Figure 2. System model structure's with first stage synchronization

For multi-user UWB TH-PAM systems, a correlation between the two adjacent symbol-long segments $r(t - kT_s)$ and $r(t - (k - 1)T_s)$ is achieved. Let $x(k;\tau)$ the value of this correlation $\forall k \in [1, +\infty)$ and $\tau \in [0, T_s)$,

$$x(k;\tau) = \sum_{u=0}^{N_u - 1} \int_0^{T_s} r(t - kT_s) r(t - (k - 1)T_s) \, dt \qquad (6)$$

After development and calculation, we find the above equation as,

$$x(k;\tau) = x_0(k;\tau) + \sum_{u \neq 0} s_u(k - 1)\big[ s_u(k)\varepsilon_{u,B}(\tilde{\tau}_u) + s_u(k - 2)\varepsilon_{u,A}(\tilde{\tau}_u)\big] + \xi(k;\tau) \qquad (7)$$

where $s_u(k)$'s are zero-mean information symbols emitted by the $(u \neq 0)$th user, $\varepsilon_{u,A}(\tilde{\tau}_u) := \varepsilon_u \int_{T_s - \tilde{\tau}_u}^{T_s} p_{u,R}^2(t)dt$, $\varepsilon_{u,B}(\tilde{\tau}_u) := \varepsilon_u \int_0^{T_s - \tilde{\tau}_u} p_{u,R}^2(t)dt$, $\tilde{\tau}_u := |\tau_u - \tau|_{T_s}$ and $\xi(k;\tau)$ corresponds to the superposition of three noise terms [10] and can be approximated as an additive white Gaussian noise (AWGN) with zero mean and $\sigma_\xi^2$ power.

Practically, the mean square of $x^2(k;\tau)$ is approximated from the average of different values $x^2(k;\tau)$ for k ranging from 0 to M–1 acquired during an observation interval duration's $MT_s$. In the multi-user case, for the two synchronization modes NDA and DA, the TDT algorithm is given as follows,

$$\hat{\tau}_u = argmax_{\tau \in [0,T_s]} E\{x^2(k;\tau)\}$$

$$x(M;\tau_u) = \frac{1}{M} \sum_{m=0}^{M-1} (x(k;\tau))^2 \qquad (8)$$

As a second stage, we propose a new fine synchronization approach with low complexity in order to have a more accurate estimate of the exact time synchronization. In this stage, we analyze and develop a fine synchronization algorithm that will give a better estimate of the time delay. The second stage synchronization description's is given in Fig 3.

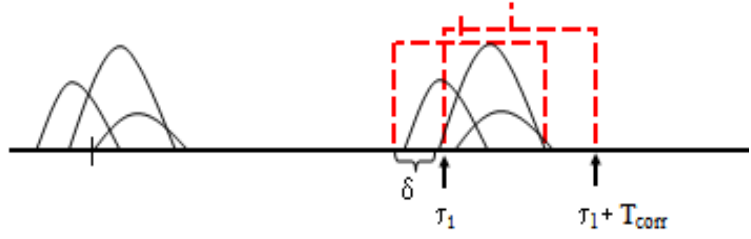

Figure 3. Description of second synchronization stage

This floor realizes a fine estimation of the frame beginning, after a blind research in the first. The idea which is based this floor is extremely simple. The proposition is to scan the time period $[\tau_1 - T_{corr}, \tau_1 + T_{corr}]$ through a step noted δ by making integration between the received signal and its replica shifted by $T_f$ on a window with a width $T_{corr}$. $\tau_1$ is the estimate delay found after the first synchronization stage. We denote the integration window output for the n$^{th}$ step nδ as follows

$$Z_n = \sum_{k=0}^{K-1} \left| \int_{t_1+n\delta}^{\tau_1+n\delta+T_s} r(t - kT_s) r(t - (k+1)T_s) dt \right| \quad (9)$$

where $n = -N + 1..0..N - 1$, $N = \lfloor T_{corr}/\delta \rfloor$ and K is the frames number considered for enhancing the decision taken at the first stage. The value of n which maximizes $Z_n$ gives the exact moment of pulse beginning noted $\tau_2 = \tau_1 + n_{opt}\delta$. Consequently, the fine synchronization is achieved. Finally, note that this proposed timing acquisition approach will be developed in both NDA and DA modes. In the next Section, we will deduce for multi-user environments in what mode this approach gives us better results compared to those given by the original approach TDT in terms of mean square error (MSE) and especially acquisition probability.

## 4. SIMULATION RESULTS AND COMPARISONS

In this section, we will evaluate the performance of our timing acquisition approach in multi-user environments with simulations. The UWB pulse is the second derivative of the Gaussian function with unit energy and duration $T_p \approx 0.8ns$. Simulations are achieved in the IEEE 802.15.3a channel model CM1 [15]. The sampling frequency chosen in the simulations is $f_c$ = 50 GHz. Each symbol contains $N_f$ = 32 frames each with duration $T_f$ = 35 ns. We used a random TH code uniformly distributed over $[0, N_c - 1]$, with $N_c$ = 35 and $T_c$ = 1.0 ns. The width integration window value's $T_{corr}$ is 4 ns. The performance of our approach is tested for various values of M with the presence of two interfering users. The two interfering users are asynchronous relative to the desired user, and are sending information symbols with 5 and 10 dB less SNR than the desired one.

In Figs. 4-5, we compare the multi-user performances in terms of mean square error (MSE) of both original TDT and fine synchronization approaches for different values of M [4]. From the simulation results, we note that increasing the duration of the observation interval M leads to improved performance for both NDA and DA modes. In comparison with the original TDT approach, we show also that the new timing acquisition approach greatly outperforms the NDA mode and offers a slight improvement in DA mode.

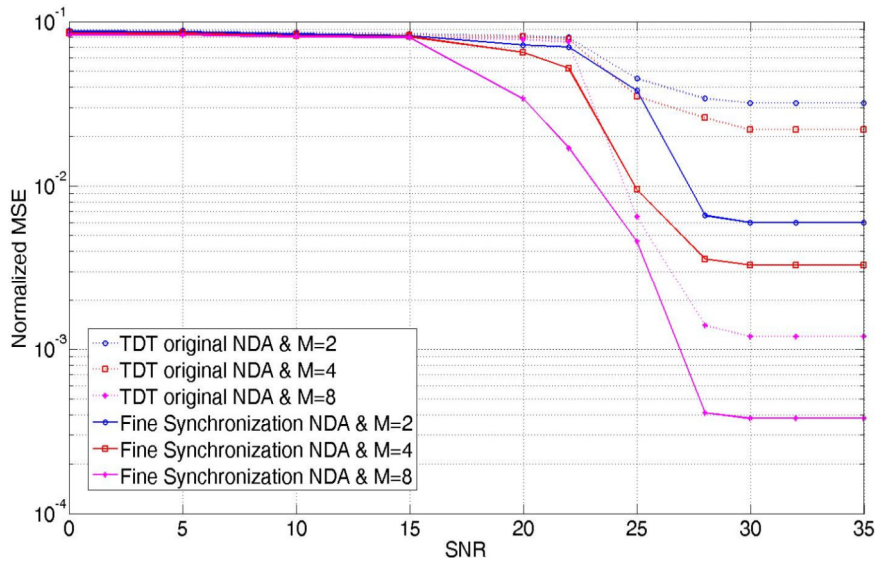

Figure 4. MSE Performances comparison in NDA mode with multi-user environments

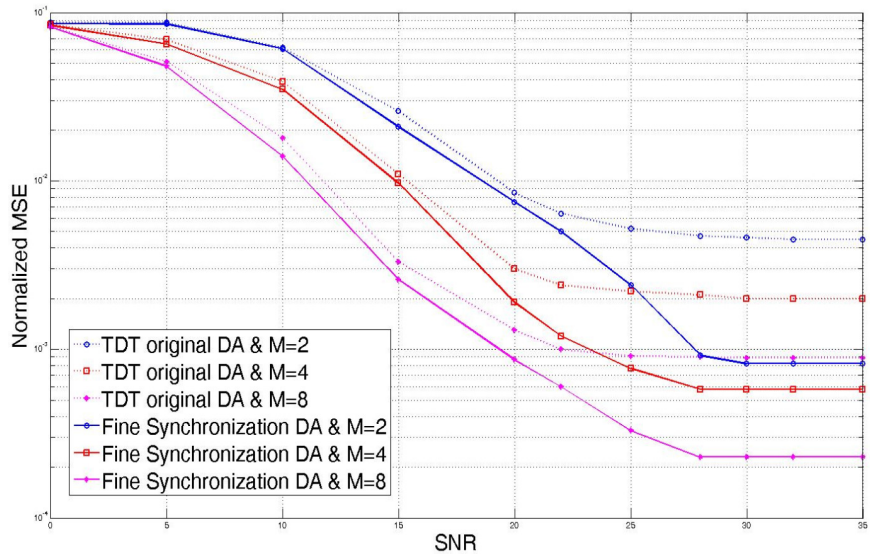

Figure 5. MSE Performances comparison in DA mode with multi-user environments

In Figs. 6-7, we compare the timing acquisition probability of the proposed synchronizer with the original TDT algorithm. From the simulation results, we deduce that our novel synchronizer can ameliorate the original NDA TDT algorithm and realize a slight improvement performance's to the DA TDT algorithm. This performance amelioration is enabled thanks to the contribution of fine synchronization approach introduced in second stage which can further improve the timing offset found in first stage (coarse synchronization approach: TDT). Unfortunately, this performance amelioration is permitted at the price of higher computation complexity and consequently time lost.

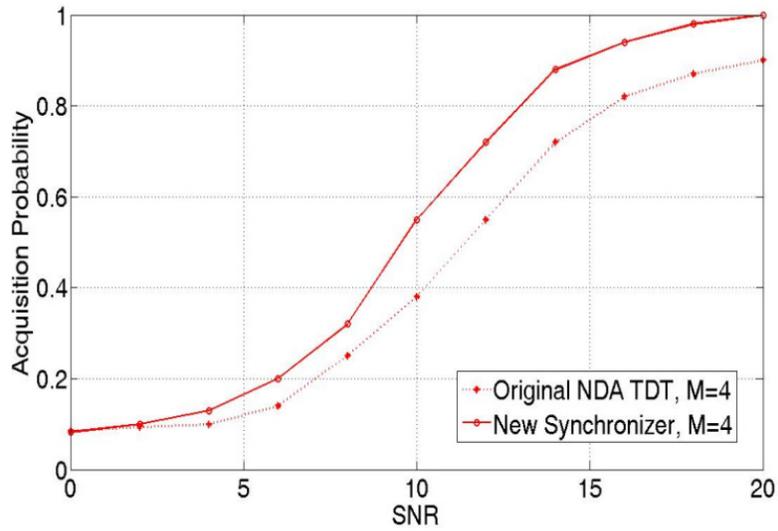

Figure 6. Multi-user acquisition probability comparison in NDA mode: proposed synchronizer vs. original TDT

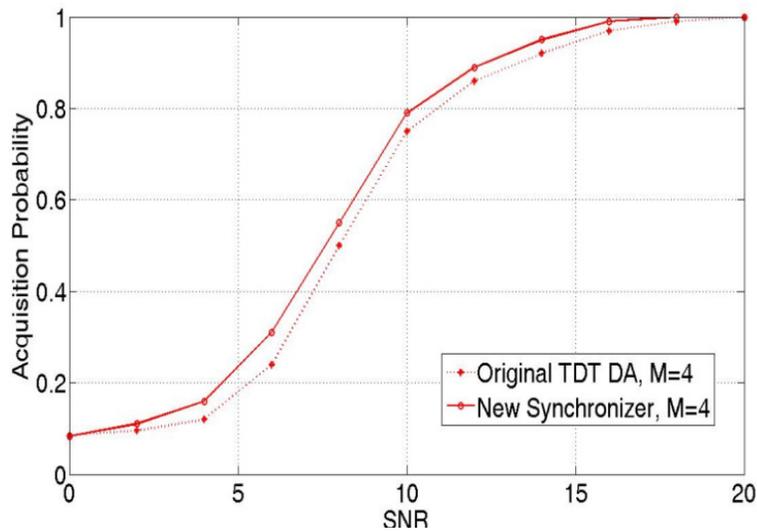

Figure 7. Multi-user acquisition probability comparison in DA mode: proposed synchronizer vs. original TDT

## 5. CONCLUSIONS

In this paper, we establish a new timing acquisition approach based on the timing with dirty templates (TDT) introduced in first stage of our synchronization algorithm for UWB radio system. With the fine synchronization approach introduced in second stage, we realize a fine estimation of the frame beginning. The simulation results show that even without training symbols, our novel synchronizer can enable a better performance (in terms of mean square error and acquisition probability) than the original TDT especially in NDA mode and offers a slight improvement in DA mode.

**Moez HIZEM**

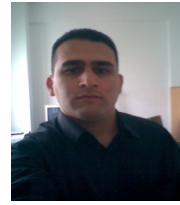

He received the M.S degree in Electronics in 2004 from the Science Faculty of Tunis (FST), Tunisia, the M.Sc. degree in Telecommunications in 2006 and the Ph.D. Degree in Telecommunications in 2011 from the National Engineer School of Tunis (ENIT), Tunisia. Since September 2011, he was an associate professor in the High Institute of Applied Sciences and Technologies (ISSAT), and has taught courses in digital Transmission, mobile and satellite Communications. He is currently working toward the Hd.R. degrees in Telecommunication at the High School of Telecommunication of Tunis (SUP'com) in the Laboratory research of Innovation of Communication and Cooperative Mobiles (Innov'COM), Tunisia. His current research interests include Wireless systems, Modulation formats, Ultra Wideband Systems, and Cooperative and Cognitive radio.

**Ridha BOUALLEGUE**

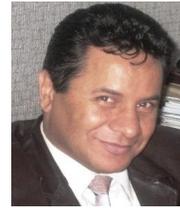

He received the Ph.D degrees in electronic engineering from the National Engineering School of Tunis. In Mars 2003, he received the Hd.R degrees in multiuser detection in wireless communications. From September 1990 he was a graduate Professor in the higher school of communications of Tunis (SUP'COM), he has taught courses in communications and electronics. From 2005 to 2008, he was the Director of the National engineering school of Sousse. In 2006, he was a member of the national committee of science technology. Since 2005, he was the Innov'COM laboratory research in telecommunication Director's at SUP'COM.
From 2005, he served as a member of the scientific committee of validation of thesis and Hd.R in the higher engineering school of Tunis. His current research interests include wireless and mobile communications, OFDM, space-time processing for wireless systems, multiuser detection, wireless multimedia communications, and CDMA systems.